\documentclass[aps,amsmath,amssymb,nofootinbib,preprint numbers,twocolumn]{revtex4} 
\usepackage{graphicx} 
\usepackage{amsmath}
\usepackage{amsfonts,amsbsy}
\usepackage{amssymb}
\usepackage{physics}
\usepackage[caption = false]{subfig}

\allowdisplaybreaks

\usepackage{xcolor}
\definecolor{lcolor}{rgb}{0.5,0,0}
\definecolor{citcolor}{rgb}{0,0.3,0.0}
\usepackage[breaklinks,colorlinks,urlcolor=blue,citecolor=citcolor,linkcolor=lcolor]{hyperref}

\usepackage{mciteplus}

\def\gsim{ \,\, \vcenter{\hbox{$\buildrel{\displaystyle >}\over\sim$}}
 \,\,}

\def\be{\begin{equation}}
\def\ee{\end{equation}}
\def\bea{\begin{eqnarray}}
\def\eea{\end{eqnarray}}

\newcommand{\as}{\alpha_{\mathrm{s}}}

\newcommand{\aem}{\alpha_\mathrm{em}}
\newcommand{\rt}{\mathbf{r}}
\newcommand{\bt}{\mathbf{b}}
\newcommand{\Kt}{\mathbf{K}}
\newcommand{\qt}{\mathbf{q}}
\newcommand{\kt}{\mathbf{k}}
\newcommand{\qso}{Q_{s,0}}

\newcommand{\jpsi}{$\mathrm{J}/\psi$\ }

\begin{document}

\title{High-energy dipole scattering amplitude 
from evolution of   low-energy proton light-cone wave functions}

\preprint{HIP-2023-6/TH }

\author{Adrian Dumitru}
\email{adrian.dumitru@baruch.cuny.edu}
\affiliation{Department of Natural Sciences, Baruch College, CUNY,
17 Lexington Avenue, New York, NY 10010, USA}
\affiliation{The Graduate School and University Center, The City University
  of New York, 365 Fifth Avenue, New York, NY 10016, USA}

\author{Heikki Mäntysaari}
\email{heikki.mantysaari@jyu.fi}
\affiliation{
Department of Physics, University of Jyväskylä,  P.O. Box 35, 40014 University of Jyväskylä, Finland
}
\affiliation{
Helsinki Institute of Physics, P.O. Box 64, 00014 University of Helsinki, Finland
}

\author{Risto Paatelainen} \email{risto.paatelainen@helsinki.fi}
\affiliation{Helsinki Institute of Physics and Department of Physics, FI-00014 University of Helsinki, Finland}

\begin{abstract}
The forward scattering amplitude of a small dipole at high
energies is given in the mean field approximation by the
Balitsky-Kovchegov (BK) evolution equation. 
It requires an
initial condition $N(r; x_0)$ describing the scattering of a dipole with size $r$ off the target that is probed at momentum fraction $x_0$. Rather than using ad hoc
parameterizations tuned to high-energy data at
$x\ll x_0$, here we attempt to construct an initial scattering
amplitude that is consistent with low-energy, large-$x$ properties of the
proton.  We start from a non-perturbative three quark light-cone
model wave function from the literature. We add ${\cal O}(g)$
corrections due to the emission of a gluon, and ${\cal O}(g^2)$
virtual corrections due to the exchange of a gluon, computed in
light-cone perturbation theory with exact kinematics. We provide
numerical data as well as analytic parameterizations of the
resulting $N(r; x_0)$ for $x_0=0.01 - 0.05$.  Solving the BK equation in
the leading logarithmic (LL) approximation towards lower $x$, we
obtain a fair description of the charm cross section in deeply
inelastic scattering measured at HERA by fitting one
parameter, the coupling constant $\alpha_s\simeq 0.2$.
However, without the option to tune the initial amplitude at $x_0$,
the fit of the high precision data results in $\chi^2/N_\text{dof} =
2.3$ at $N_\text{dof} =38$, 
providing clear statistical evidence for the need of 
{\em systematic} improvement
e.g.\ of the photon wave function, evolution equation, and
initial condition.
\end{abstract}

\maketitle

\section{Introduction}

In Deep Inelastic Scattering (DIS) a pointlike virtual photon probes the rich QCD dynamics taking place inside the proton or a nucleus. At high energies, where the small Bjorken-$x$ part of the target wave function is probed, one observes very large gluon densities~\cite{H1:2015ubc}. When the gluon densities become of the same order as inverse coupling, non-linear QCD dynamics start to dominate and multiple scattering effects are important~\cite{Kovchegov:2012mbw}. In the high-energy limit, the scattering process is most conveniently described in the dipole picture in a frame where the virtual photon has a large momentum~\cite{Mueller:1994jq}, and its partonic Fock states, such as $|q\bar q\rangle$ at leading order (LO), have a long lifetime as they scatter from the color field of the target.

Describing the QCD dynamics in this high-density domain is natural in the Color Glass Condensate~\cite{Gelis:2010nm} framework. Here the center-of-mass energy or Bjorken-$x$ dependence of various observables (and as such the target structure) is described in the large-$N_\mathrm{c}$ limit by the perturbative Balitsky-Kovchegov (BK) renormalization group equation~\cite{Kovchegov:1999yj,Balitsky:1995ub}. It describes how the dipole-target scattering amplitude, which contains information about the target structure, changes with increasing energy. The dipole amplitude (a correlator of two Wilson lines) is actually a convenient degree of freedom at high energies: all cross sections computed at high energy in the CGC framework are expressed in terms of the dipole amplitude or higher-point correlators which can be written, in a Gaussian approximation, in terms of the two-point dipole amplitude~\cite{Dominguez:2011wm}. 

The initial condition for the dipole-proton scattering amplitude depends on non-perturbative properties of the proton. A typical approach in the field has been to assume an intuitive functional form at an initial $x_0 \ll 1$ and fit various unknown parameters to the HERA total cross section data; see, e.g., Refs.~\cite{Albacete:2009fh,Albacete:2010sy,Lappi:2013zma} where a very good description of small-$x$ HERA data is obtained at leading order, resumming powers of $\as \ln 1/x$ via BK evolution with running coupling corrections~\cite{Balitsky:2006wa}. Recent developments to full NLO accuracy have also allowed for a simultaneous description of total and heavy quark production data~\cite{Beuf:2020dxl,Hanninen:2022gje}.
The drawback of this approach is that one is sensitive to the assumed functional form of the initial dipole amplitude and that the model parameters need to be re-fitted if the evolution is initialized at different $x_0$. Furthermore, there is no relation to the low energy (
or ``large-$x$") proton structure.

In this work, we take a complementary approach aiming to \emph{compute} the initial dipole-proton scattering amplitude at moderate $x_0$. As we will discuss in more detail next, the necessary non-perturbative input consists in a proton valence quark wave function that is constrained by low-energy data. The $x_0$-dependent initial condition is then obtained by computing the dipole-target scattering amplitude including one perturbative gluon emission in the target, with the gluon longitudinal momentum fraction regulated by $x_0$~\cite{Dumitru:2020gla,Dumitru:2021tvw}. The advantages of this approach are that we do not assume an ad-hoc functional form of the scattering amplitude and that the initial condition can be computed and the BK evolution initialized at any (moderate) $x_0$ without a need to perform new fits. Also, this approach largely eliminates the freedom of tuning initial conditions in order to optimally match the
evolution equation to the small-$x$ data. This may reveal quantitative evidence for the need for improvements beyond leading-log, or
even running coupling BK evolution.

Finally, we would like to point out that light-cone Hamiltonian calculations of wave functions have been employed previously to
set initial conditions for QCD scale evolution to high virtuality $Q^2$, 
in order to describe DIS in the ''dipole
approach" using correlators of eikonal Wilson lines as degrees of 
freedom~\cite{Hautmann:1999ui,Hautmann:2000pw,Hautmann:2006xc,Hautmann:2007cx}.
Our approach is similar in spirit although
here the goal is to determine initial conditions for evolution to small $x$.

\section{Dipole-proton scattering at moderate \texorpdfstring{$x$}{x}}
\label{sec:dipole_lcpt}

We first provide an overview of our approach to the light-cone structure of the proton. We employ
a truncated Fock space description which starts with a three quark state. The
corresponding Fock space amplitude (wave function) $\Psi_\mathrm{qqq}$ corresponds to a
non-perturbative solution of the QCD light-front Hamiltonian. 
To date, exact solutions for the light-cone wave functions are not
available. In the future, lattice gauge theory may provide
numerical solutions for moderate parton momentum fractions $x_i$ and
transverse momenta $\vec k_i$ via a large momentum expansion of
equal-time Euclidean correlation functions in instant
quantization~\cite{Ji:2020ect,Ji:2021znw}; see ref.~\cite{Chu:2023jia} for a recent
lattice computation of the wave function of the leading $q\overline q$ state of the pion.
Also, the MAP collaboration~\cite{Pasquini:2023aaf} has recently extracted the wave functions of the first four
Fock states of the pion from fits to its parton distribution functions and
electromagnetic form factor.

Here, we rely on
solutions of effective light-cone Hamiltonians for guidance on the
low energy and low virtuality $Q^2$ 
structure of the proton. Specifically, we shall employ
the HO wave function of Refs.~\cite{Schlumpf:1992vq,Brodsky:1994fz}.
In these references, the authors fixed the parameters of the three quark
wave function to the proton ``radius", or Dirac form factor at $Q^2\to0$,
to the anomalous magnetic moments of the proton and neutron, and to the
axial vector coupling $g_A$. The wave function also matches
reasonably well the empirical knowledge of the longitudinal and
transverse momentum distribution of single quarks in the valence quark
regime. Finally, the wave function of Refs.~\cite{Schlumpf:1992vq,Brodsky:1994fz}
also provide predictions for quark {\em momentum correlations}.

At next-to-leading order (NLO) in the Fock expansion we add the three quarks and one gluon state
with amplitude $\Psi_\mathrm{qqqg}$, as well as
the virtual corrections to $\Psi_\mathrm{qqq}$ due to the
exchange of a gluon by two quarks in the proton. These corrections are
obtained via light-cone perturbation theory calculations~\cite{Dumitru:2020gla,Dumitru:2021tvw}.
The presence or exchange of the gluon extends the range of parton light-cone
momentum fractions to lower $x$, and pushes their
transverse momenta into the perturbative regime. It also affects
their momentum correlations.

The central element of our analysis is the (imaginary part of the)
eikonal scattering amplitude $N$ of
a small dipole of transverse size $\rt$. The real part of $N$
corresponds to two-gluon exchange,
\begin{multline}
\label{eq:dipole_from_lcpt}
  N(\rt,\bt) = -g^4 C_\mathrm{F}
  \int \frac{\dd^2 \Kt \, \dd^2 \mathbf{q}}{(2\pi)^4}
  \frac{\cos\left(\bt \cdot \Kt\right)}{(\mathbf{q} - \frac{1}{2}\Kt)^2\,\,
    (\mathbf{q} + \frac{1}{2} \Kt)^2} \\
  \times \left( \cos(\rt \cdot \mathbf{q}) -
  \cos\left(\frac{\rt \cdot \Kt}{2} \right) \!\! \right)
   G_2\left(\mathbf{q} -\frac{1}{2}\Kt, -\mathbf{q} - \frac{1}{2} \Kt\right). 
\end{multline}
Here $\Kt$ is the momentum transfer which is Fourier conjugate to the
impact parameter $\bt$. As explained below, we will eventually average
$N(\rt,\bt)$ over a suitable range of impact parameters.
We emphasize that the expression above accounts only for a single,
perturbative two-gluon exchange (see its derivation in Ref.~\cite{Dumitru:2018vpr}),
it does not resum the
Glauber-Mueller multiple scattering series. This restricts its
applicability to the regime of weak scattering, $N(\rt,\bt) \ll 1$.
Furthermore, $N(\rt,\bt)$ actually acquires an imaginary part due
to the perturbative exchange of three gluons; its magnitude has been shown to
be much smaller than its real part~\cite{Dumitru:2021tqp,Dumitru:2022ooz}, 
and in practice it is of interest
only for processes involving $C$-conjugation odd exchanges~\cite{Dumitru:2019qec}. For the
present purposes, it can be neglected.

The coupling of the two static gluons to the proton is described in terms
of the color charge density correlator
\begin{equation}
\label{eq:g2def}
\langle \rho^a(\qt_1)\, \rho^b(\qt_2) \rangle \equiv \delta^{ab}\, g^2
G_2(\qt_1,\qt_2).
\end{equation}
The color charge density operator corresponds to the
light-cone plus component of the color current
on the $x^+=0$ light front, integrated over $x^-$, $\rho^a(\qt) \equiv J^{+a}(\qt)$,
when the proton carries positive $P_z$.
Eqs.~(\ref{eq:dipole_from_lcpt},\ref{eq:g2def}) correspond to the leading twist
contribution to the matrix element of the dipole operator in the proton.
Dozens of diagrams contribute to this correlator at NLO, their
explicit expressions are listed in Ref.~\cite{Dumitru:2020gla}.
We point out that $G_2(\qt_1,\qt_2)$ satisfies a Ward identity due
to the color neutrality of the proton; it vanishes when either
$q_1^2$ or $q_2^2\to 0$ so that $N(\rt,\bt)$ in 
Eq.~(\ref{eq:dipole_from_lcpt}) is free of IR divergences.
However, $G_2$ does exhibit a collinear singularity which is
regularized by assigning a mass to the quarks in the 
light-cone energy denominators for the $q\to qg$ and $qg\to q$ vertices;
see Ref.~\cite{Dumitru:2020gla} for details.
All the results presented here were obtained with $m_\text{coll} = 0.2\,\mathrm{GeV}$. This is
consistent with the quark mass and transverse momentum scales which appear in the
non-perturbative three quark wave function of Refs.~\cite{Schlumpf:1992vq,Brodsky:1994fz}.
The color charge correlator also exhibits a soft singularity when the
light-cone momentum fraction $x_g$ of the gluon goes to zero. This is
regularized with a cutoff $x$ on $x_g$, and the resummation of yet softer
gluons will be performed through the BK equation. 
Note that at $x=0.1$ the NLO contribution to $G_2$ truly is
a reasonably small ${\cal O}(g^2)$ perturbative correction~\cite{Dumitru:2021tvw}. However,
by $x=0.01$ its magnitude grows to essentially ${\cal O}(1)$, a
leading-log correction. Hence, at such $x$ resummation is required and it is justified to use the computed dipole as an initial condition for the leading order BK evolution.

We recall, also,
that at the given order ultraviolet divergences cancel~\cite{Dumitru:2020gla},
so that $G_2$ is independent of the renormalization scale,
and that the coupling does not run.
Lastly, let us mention that the angular dependence of the correlator $G_2$, as well as the
dependence of its Fourier transform on impact parameter,
has been analyzed numerically in detail in Ref.~\cite{Dumitru:2021tvw}.

\section{Small-\texorpdfstring{$x$}{x} evolution of the proton light-cone wave function}

In order to obtain an initial condition for $\bt$-independent BK evolution\footnote{We limit ourselves to the $\bt$-independent evolution in order to avoid the need to effectively model confinement scale effects which has been attempted e.g. in Refs.~\cite{Berger:2011ew,Mantysaari:2018zdd}.} we average the dipole-target scattering amplitude obtained from Eq.~\eqref{eq:dipole_from_lcpt} over the impact parameter $\bt$,
\begin{equation}
\label{eq:ic_bavg}
    N(r,x_0) = \frac{1}{S_T} \int^{b_\mathrm{max}} \dd[2]\bt \, N(\rt,\bt,x_0).
\end{equation}
Throughout this work, we denote the magnitudes of the transverse vectors as $b=|\bt|$ and $r=|\rt|$. 
The resulting amplitude is dominated by perturbative contributions when the dipole size $r$ is small. In this region there is a small $\cos(2\phi)$ dependence on the angle $\phi$ between $\rt$ and $\bt$~\cite{Dumitru:2021tvw} which vanishes when we integrate over $\bt$. Here $S_T$ is the proton transverse area. Inclusive cross sections considered in this work are not sensitive to the actual shape of the target but only to the total transverse size. The proton geometry is most directly probed in exclusive vector meson production process where the total momentum transfer $\Kt$ which is  Fourier conjugate to the impact parameter is measurable. Parametrizing the \jpsi production cross section in HERA kinematics as $e^{-B_D \Kt^2}$ one obtains $B_D= 4\,\mathrm{GeV}^{-2}$~\cite{H1:2005dtp}. Assuming a Gaussian impact parameter profile for the proton, this corresponds to a two-dimensional root-mean-square radius $b_\mathrm{Gaussian}=\sqrt{2B}\approx 0.56\,\mathrm{fm}$ and a proton area $S_T=2\pi B$. On the other hand, if we assume a step function (hard sphere) profile for the proton, the same diffractive slope is obtained when the proton radius is $b_\mathrm{Hard\, sphere}=2\sqrt{B}\approx 0.79\,\mathrm{fm}$, which corresponds to $S_T=4\pi B$.

 Although exclusive vector meson data favors the Gaussian density profile over the hard sphere one (see e.g.~\cite{Kowalski:2006hc}), the current data does not constrain the proton shape precisely.
 We also note that if the $\bt$-dependent dipole amplitude from Eq.~\eqref{eq:dipole_from_lcpt} is directly used to compute exclusive \jpsi production cross section, the resulting spectra differs from the Gaussian profile case only in the region where there are no experimental constraints~\cite{Dumitru:2021hjm}. 
 In this work the results shown below by default  correspond to the Gaussian density profile (with $b_\text{max}=b_\mathrm{Gaussian}$) unless otherwise stated, but we also study the dependence on the $b_\text{max}$ cut by using a step function profile with $b_\text{max}=b_\mathrm{Hard\ sphere}=\sqrt{2}b_\mathrm{Gaussian}$.

The proton transverse area $S_T$ has also been extracted by fitting a parameterized initial condition for the BK evolution equation to the HERA structure function data. Leading order analyses~\cite{Albacete:2010sy,Lappi:2013zma} typically obtain $S_T\sim 16\,\mathrm{mb}$. In recent fits at NLO accuracy~\cite{Beuf:2020dxl,Hanninen:2022gje} proton areas $S_T \sim 10\dots 20 \,\mathrm{mb}$ were obtained depending on the details of the analysis setup. We test this uncertainty in the proton small-$x$ transverse profile by showing some results for both the Gaussian and hard sphere profiles with transverse areas $9.8\,\mathrm{mb}$ and $19.6\,\mathrm{mb}$, respectively.

Before performing the impact parameter average  we first study the impact parameter  profile from the NLO light-cone wave function seen by a perturbative probe:
\begin{equation}
\label{eq:effective_density_profile}
    T(\bt) = C \int^{r_\text{max}}  \dd[2]\rt \, N(\rt,\bt,x).
\end{equation}
The normalization condition $\int^{b_\mathrm{max}} \dd[2]{\bt} T(\bt)=1$ is used to fix the constant $C$. We will refer to $T(\bt)$ as the
transverse "density" profile to match standard terminology from the literature.
As the dipole amplitude is a rapidly increasing function of the dipole size $r$, this integral is dominated by dipoles of size $r\sim r_\text{max}$, as long as $r_\text{max}$ is in the perturbative domain.

\begin{figure}
     \centering
         \includegraphics[width=\columnwidth]{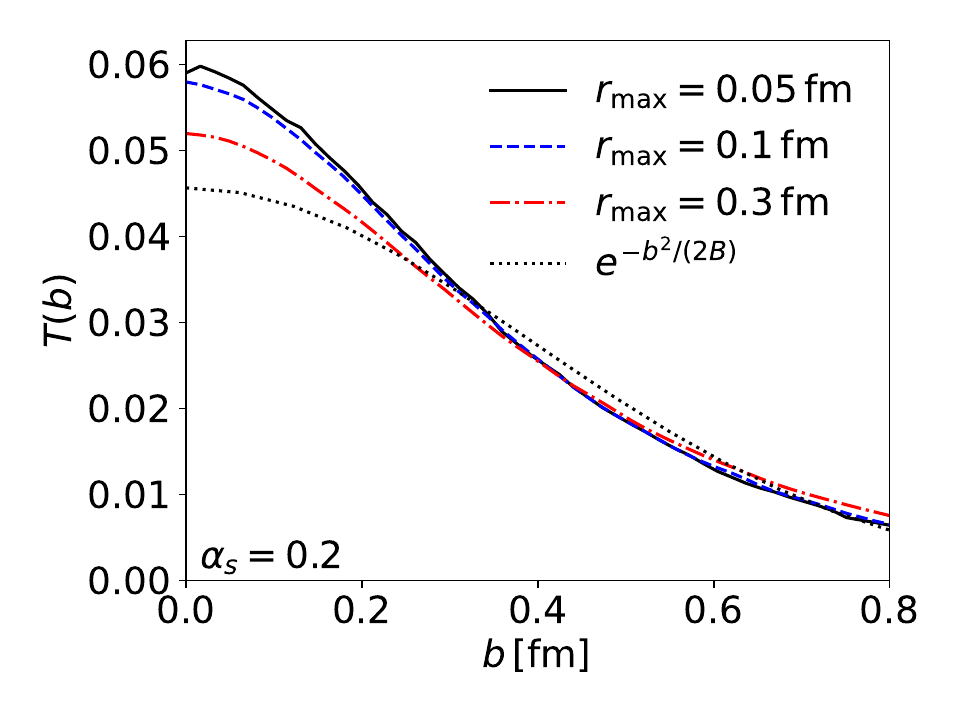}
         \label{fig:}
     \vspace*{-0.5cm}
     \caption{Effective normalized proton density profiles at $x=0.01$ extracted from $N(\rt,\bt)$ with $b_\text{max}=0.8\,\mathrm{fm}$ up to NLO in the Fock expansion
     of the light-cone wave function. }
     \label{fig:densityprofile}
\end{figure}
The extracted density profiles up to $b_\mathrm{max}=0.8\,\mathrm{fm}$ for different $r_\text{max}$ are shown in Fig.~\ref{fig:densityprofile}. 
For reference, a Gaussian profile, as used e.g.\ in the
popular IPsat parametrization~\cite{Kowalski:2003hm} for the dipole amplitude with the slope $B=4\,\mathrm{GeV}^{-2}$, is also shown.
We observe a similar transverse profile except for very central $b\lesssim 0.2$~fm where the computed profile is more steeply falling. This region can only be probed
at high momentum transfer $|t|\gsim 1$~GeV$^2$~\cite{Dumitru:2021hjm}, which is not covered in the  currently available coherent vector meson production data.
The high-$b$ tails of $T(\bt)$ resulting from the LCPT one gluon emission
corrections are exponential rather than Gaussian. However, in all we
conclude that for the present purposes
the Gaussian profile used to match $S_T$ to $b_\mathrm{max}$ is a reasonable approximation.

\begin{figure}
    \centering
    \includegraphics[width=\columnwidth]{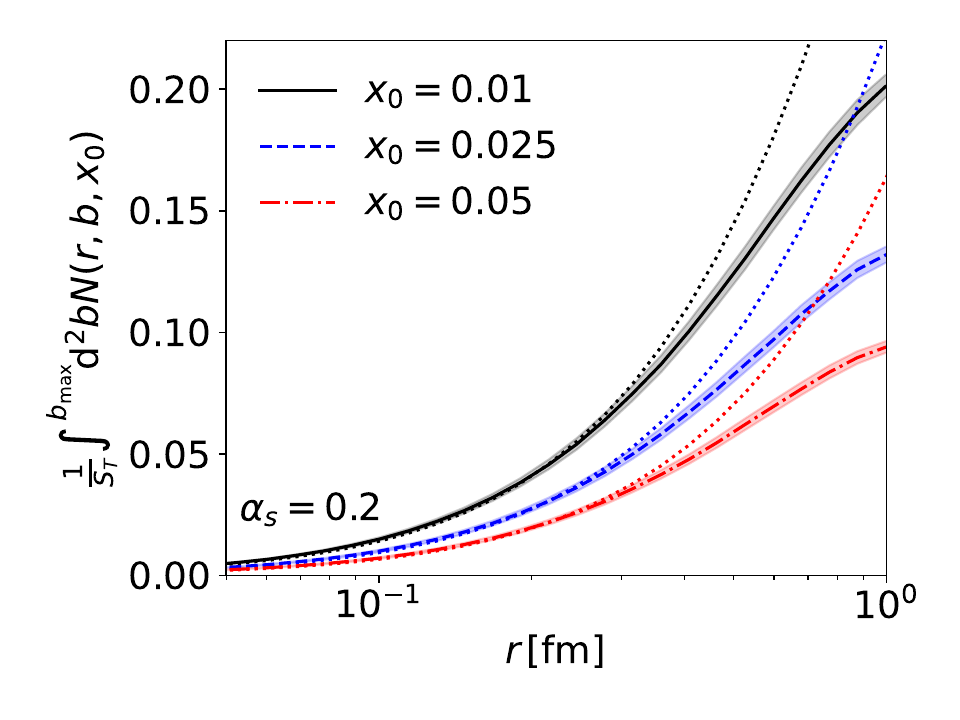}
    \caption{Impact-parameter averaged dipole as a function of dipole size $r$ at two different momentum fractions $x$. 
    The bands correspond to varying the proton shape parameter $B$ by $10\%$. The dotted lines show best fits to the central values with the modified MV model parametrization of Eq.~\eqref{eq:mv}.  
   }
    \label{fig:dipole_bavg_x0}
\end{figure}

\begin{figure}
    \centering
    \includegraphics[width=\columnwidth]{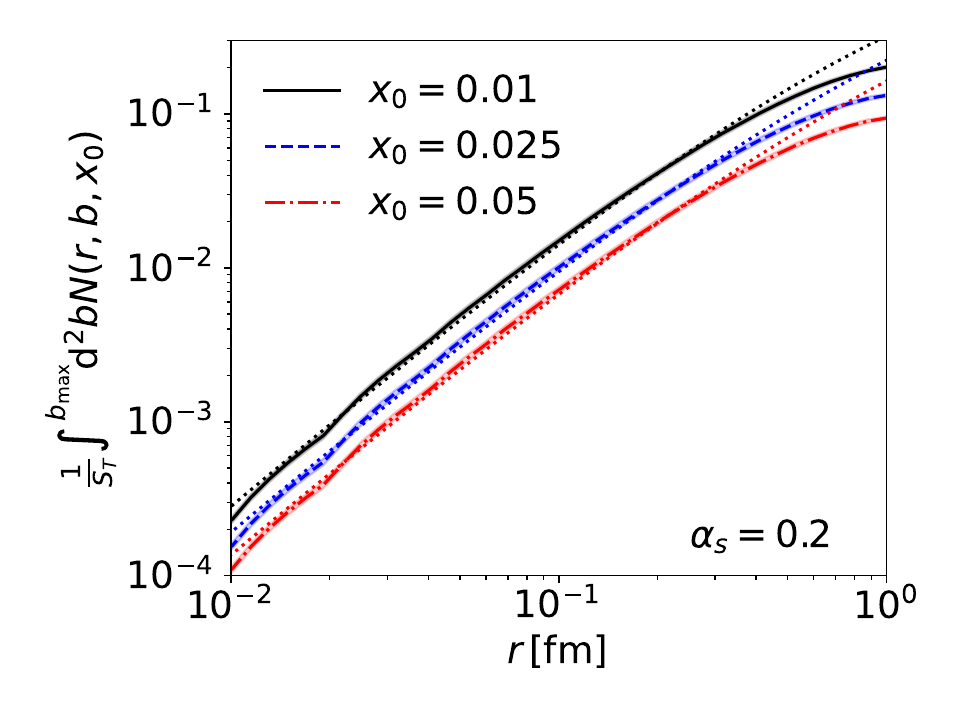}
    \caption{Same as Fig.~\ref{fig:dipole_bavg_x0} but on a double logarithmic scale in order to better exhibit the behavior at small $r$. 
   }
    \label{fig:dipole_bavg_x0_log}
\end{figure}

The $\bt$-averaged dipole amplitudes (using a Gaussian profile) are shown in Fig.~\ref{fig:dipole_bavg_x0} (linear scale) and Fig.~\ref{fig:dipole_bavg_x0_log} (logarithmic scale)  at $x_0=0.05, x_0=0.025$ and $x_0=0.01$.  Here we also show the dependence on the diffractive slope $B$: the bands correspond to varying $B$ by $\pm10$\% which changes both  $S_T$ and $b_\text{max}$. The results depend weakly on this cut especially in the perturbative small-$r$ domain. The dipole amplitude increases with $r$, approximately proportional to $r^2$, as expected. For $r\gsim 0.4$~fm
the color neutrality of the proton, and the fact that the dipole scatters from a target of finite transverse extent, begin to
slow the growth of $N(r)$;  a model that does not account for the finite size of the proton in impact parameter space
would attribute this to power corrections. Finally,
when the size of the dipole becomes comparable to that of the target the amplitude is found to decrease again (not shown)
as the end points
of the dipole essentially ``miss'' the target. However, we emphasize that this  behavior occurs at large $r\sim \mathrm{few}\,\mathrm{fm}$ where in any case the perturbative calculation of the scattering amplitude is not valid.

Figures~\ref{fig:dipole_bavg_x0} and~\ref{fig:dipole_bavg_x0_log}
confirm that down to $x=0.01$ scattering of small dipoles with $r$ significantly less than 1~fm remains quite weak,
at least for $\as=0.2$ which we determine below from a fit to the charm cross section in DIS. Therefore, it appears reasonable
to start small-$x$ evolution with this initial condition at $x$ in the range $0.01\dots 0.05$.

To obtain analytic parameterizations of the dipole amplitude we fit our numerical data for the $\bt$-averaged scattering
amplitude to the following expression which is inspired by the McLerran-Venugopalan (MV) model~\cite{McLerran:1993ni}:
\begin{equation}
    \label{eq:mv}
    N(r) = 1 - \exp\left[ -\frac{(r^2 \qso^2)^\gamma}{4} \ln \left( \frac{1}{r\Lambda} + e_c\cdot e\right) \right],
\end{equation}
where $\Lambda=0.241\,\mathrm{GeV}$ is a fixed infrared scale.
Such a parameterization has been used previously e.g. in Refs.~\cite{Albacete:2010sy,Lappi:2013zma,Beuf:2020dxl} to fit the initial condition for BK evolution to the HERA data.
While our fit is restricted to $r<0.5$~fm, the parameterization forces $N(r)\to 1$ in the large-$r$ region. Of course,
the behavior at large $r$ can not be trusted, and other extrapolations would be possible. It is important, however, that 
the large-$r$ extrapolation is such that the Fourier transform of $1-N(r)$ at high $k$ (and, consequently, the forward particle production cross section, for example) will be insensitive to
the assumed form.
We will also demonstrate below that perturbative observables, in our case the charm production cross section, are only sensitive to the perturbative regime of small dipoles where our calculation should apply, and not to the extrapolation to large $r$.

\begin{table}[tb]
\begin{center}
\begin{tabular}{c|c|c|c}
$x$ & $\qso^2\ [\mathrm{GeV}^2]$ & $\gamma$ & $e_c$ \\
\hline  
0.01 & $0.100^{+0.004}_{-0.004}$ & $1.001^{+0.001}_{-0.001}$ & $e^{-1}$ \\
0.025 & $0.066^{+0.003}_{-0.003}$ & $0.998^{+0.001}_{-0.001}$ & $e^{-1}$ \\
0.05 & $0.047^{+0.002}_{-0.002}$ & $0.997^{+0.001}_{-0.001}$ & $e^{-1}$
\end{tabular}
\caption{Best fit parameters to the modified MV model parameterization, Eq.~\eqref{eq:mv}, for a fit over $0.01\,\mathrm{fm}<r<0.5\,\mathrm{fm}$. The upper and lower limits are obtained by varying the proton shape parameter $B$  by $\pm10$\%. All fit results give $e_c=e^{-1}$ within numerical accuracy. 
\label{tab:mvfit}
}
\end{center}
\end{table}%

The free parameters in Eq.~\eqref{eq:mv}, $\qso^2, \gamma$ and $e_c$, are fit to the calculated dipole amplitude in the region $0.01\,\mathrm{fm}<r<0.5\,\mathrm{fm}$ (we actually fit the logarithm of the dipole in order to give equal weight to small and intermediate $r$).
The upper limit restricts to the perturbative domain, and the lower limit is imposed in order to give some weight to the
region of intermediate $r$ as well. The resulting dipole amplitudes are shown in Figs.~\ref{fig:dipole_bavg_x0} and~\ref{fig:dipole_bavg_x0_log} as dotted lines. The fit parameters are listed in Table~\ref{tab:mvfit}. In the fit we require that $e_c>e^{-1}$ in order to enforce positivity of the logarithm in Eq.~\eqref{eq:mv}, and all fit results give  $e_c=e^{-1}$ within numerical accuracy, i.e.\ they require as small an infrared cutoff as allowed.
The MV-model inspired parameterization is found to describe the dipole-proton scattering amplitude quite well, for all dipole sizes in the perturbative $r\lesssim 0.5\,\mathrm{fm}$ region. 
Here, of course, the linearized version of Eq.~\eqref{eq:mv} is sufficient, as it should be: recall that Eq.~\eqref{eq:dipole_from_lcpt} does not resum multiple scattering.

The momentum scale $\qso^2$ remains non-perturbative down to $x=0.01$; see below for an extraction of a ``saturation scale''
at lower $x$. However, it increases approximately as $\qso^2 \sim 1/x^{0.47}$.
The ``anomalous dimension'' of the dipole amplitude is $\gamma = 1$ within numerical accuracy. 
It appears reasonable to us that the initial condition for the evolution equation admits a power series expansion
in $r^2$, starting at its first power.
On the other hand, leading order fits to HERA total cross section data~\cite{Lappi:2013zma,Albacete:2010sy} require $\gamma\sim 1.1 - 1.2$ in order to obtain as slow a $Q^2$ dependence of the cross section as required by the HERA data~\cite{H1:2009pze,H1:2015ubc}; recent fits at next-to-leading order accuracy performed in Ref.~\cite{Beuf:2020dxl,Hanninen:2022gje} also prefer $\gamma \gtrsim 1$ when the heavy quark production data is included.
A problem with $\gamma>1$ is that it renders the (dipole) unintegrated gluon distribution 
function~\cite{Kovchegov:1998bi,Kharzeev:2003wz,Dominguez:2011wm} and the forward particle production cross section negative~\cite{Lappi:2013zma,Ducloue:2017mpb} in some range of transverse momentum $\kt_T$. The dipole amplitude obtained here does not display this issue.

\begin{figure}
    \centering
    \includegraphics[width=\columnwidth]{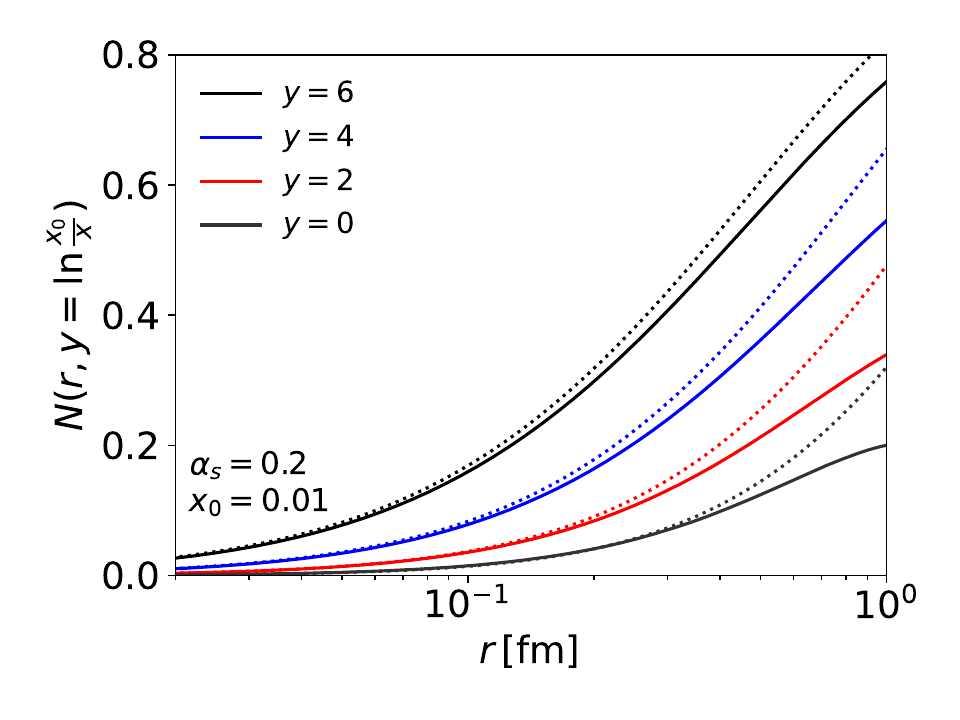}
    \caption{Leading-log BK evolution at $\as=0.2$ starting at $x_0=0.01$. From bottom to top the lines correspond to evolution
    rapidity $0, 2, 4$ and $6$. The dashed lines are obtained with the fitted MV-model like parameterization from Table~\ref{tab:mvfit} as an initial condition; the solid lines evolve the actual numerical data for $N(|\rt|,x_0)$. 
   }
    \label{fig:bkevol_dipole_maxy_10}
\end{figure}

Next we solve the leading order BK equation with fixed coupling, using the numerical data for $N(r,x_0)$ as an initial condition at $x_0=0.01$. Note that at this order in $\as$ the coupling constant does not run in the LCPT calculation of the initial condition, and consequently we also limit ourselves to the fixed coupling case here. Evolution over 6 units of rapidity is shown in Fig.~\ref{fig:bkevol_dipole_maxy_10}. For comparison, we also solve the BK equation using the modified MV-model initial condition with parameters as shown in Table~\ref{tab:mvfit}. This parameterized initial condition has a completely different behavior in the infrared region with $N(r)\to 1$ at large $r$ whereas the numerical data gives a decreasing $N(r)$ when $r$ exceeds a few fm, as already mentioned above. However, as can be seen in Fig.~\ref{fig:bkevol_dipole_maxy_10} the resulting BK-evolved dipole amplitudes are basically identical in the perturbative $r\lesssim 0.5\,\mathrm{fm}$ domain.
In fact, due to the approach to the fixed point of the BK equation~\cite{Stasto:2000er,Munier:2003vc,Munier:2003sj}, at high rapidity the difference between the scattering amplitudes evolved with the two initial conditions diminishes. This demonstrates that the BK-evolved amplitude
at small $r$ is not affected by the uncontrolled large-$r$ extrapolation of the initial condition.

One may define a saturation radius $r_s$, and a corresponding saturation momentum $Q_s=\sqrt 2/r_s$ from the condition that $N(r_s)=1-\exp(-\frac{1}{2})\simeq 0.4$. For this to be a perturbative scale
requires about 6 units of rapidity evolution, as can also be seen from Fig.~\ref{fig:bkevol_dipole_maxy_10}.  This corresponds to $x\simeq 2.5\cdot 10^{-5}$, where $r_s\simeq0.3$~fm, and $Q_s\simeq 1$~GeV.  These values are not very far from the first ``saturation model'' fit to HERA DIS
data by Golec-Biernat and W\"usthoff~\cite{Golec-Biernat:1998zce} from 25 years ago. Many more recent fits mentioned above have since confirmed that reaching the strong scattering regime with a small dipole
and a proton target requires deep evolution to rather small $x$. Also, some studies~\cite{Hautmann:2000pw} of diffractive small-$x$ scattering of a $q\bar{q}-g$ state
from the proton have indicated that the regime of "color transparency" sets in when the typical transverse distance of the gluon
from the $q\bar{q}$ is around 0.2~fm.

\begin{figure}
    \centering
    \includegraphics[width=\columnwidth]{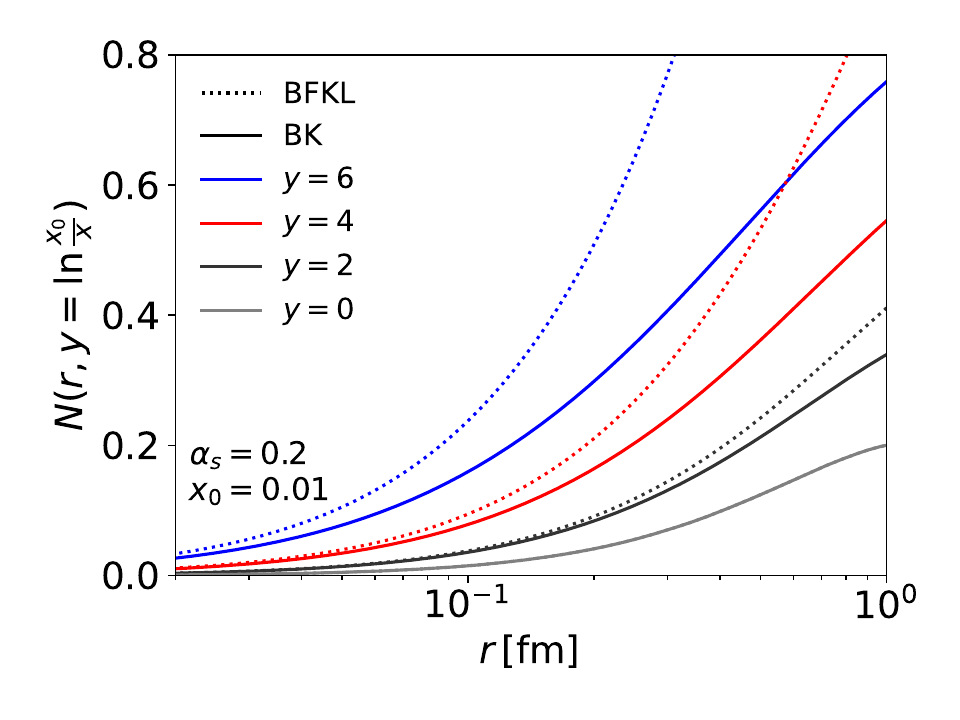}
    \caption{Leading-log BK (solid lines) vs.\ BFKL (dotted lines) evolution starting at $x_0=0.01$.   
   }
    \label{fig:bfklevol_dipole}
\end{figure}

Since scattering at $x=0.01$ is fairly weak we have also evolved our initial condition with the linear BFKL 
equation~\cite{Lipatov:1976zz,Kuraev:1977fs,Balitsky:1978ic}, see Fig.~\ref{fig:bfklevol_dipole}. 
After a few units of rapidity evolution, the linear equation begins to
violate unitarity, $N(r)\le1$, at large $r$. However, this regime of large dipoles is not under control in any case. More importantly though, at $y=2-4$ the absence of the non-linear correction begins to
affect the solution significantly even at $r$ substantially less than 1~fm. With BFKL we also noticed a greater difference between evolving the actual numerical data vs.\ the
analytic modified MV-model parametrization (not shown), which differ in their large-$r$ extrapolation. Therefore, for accurate results it appears to be rather important to evolve with the non-linear BK equation
even if one restricts to $r<1$~fm.

\begin{figure}
    \centering
    \includegraphics[width=\columnwidth]{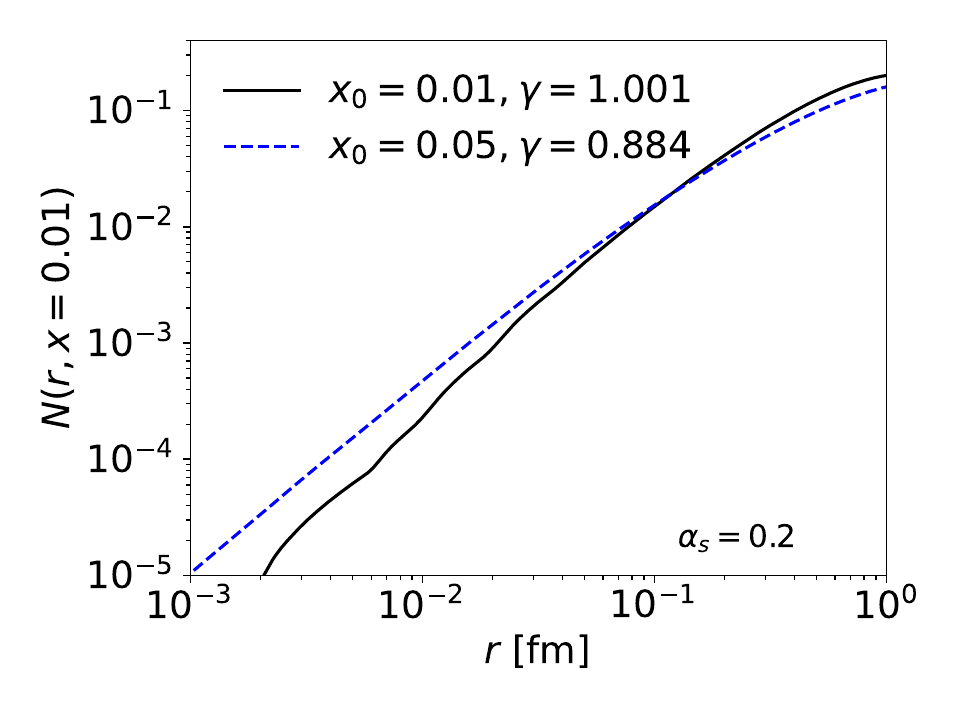}
    \caption{The dipole scattering amplitude at $x=0.01$ with and without prior BK evolution. 
    The parameter $\gamma$ denotes the resulting anomalous dimension fitted in the region $0.01\,\mathrm{fm}<r<0.5\,\mathrm{fm}$.
   }
    \label{fig:dipole_x001_x0dep}
\end{figure}

Let us finally study how the $x$ dependence obtained from the direct, fixed order NLO LCPT calculation differs to the one obtained by solving the BK equation. We note that in the LCPT calculation $x$ is an explicit cutoff for the  longitudinal momentum of the emitted gluon, and this gluon emission is calculated in exact kinematics. On the other hand, in BK evolution multiple soft gluon emissions are resummed. 
This comparison is done by calculating the dipole amplitude at $x=0.01$ directly from the LCPT using Eq.~\eqref{eq:dipole_from_lcpt}, and comparing that to the dipole amplitude obtained by solving the BK equation with the initial condition computed at $x_0=0.05$. The resulting dipole amplitudes are shown in Fig.~\ref{fig:dipole_x001_x0dep}. The most significant difference between the fixed order ${\cal O}(g^2)$ LCPT amplitude and the BK evolved dipole is that the evolution equation decreases the anomalous dimension $\gamma$ towards the asymptotic value $\gamma\sim 0.6$. On the other hand, the emission of one gluon in the direct LCPT calculation does not modify the extracted anomalous dimension, as can also be seen from Table.~\ref{tab:mvfit}. This is, of course, the expected behavior.
As already mentioned above, DIS phenomenology does not appear
to support $\gamma<1$ at $x=0.01$ or greater, so it seems reasonable to treat at least the emission of the first gluon 
with $x_g>0.01$ explicitly in fixed order light-cone perturbation theory with exact kinematics\footnote{Also, the emission of the first gluon actually increases the imaginary part due to
$C$-odd three gluon exchange~\cite{Dumitru:2022ooz}, which provides another indication that small-$x$ evolution should not be started
much before $x_0\simeq0.01$.}.

\section{Total cross section at small \texorpdfstring{$x$}{x}}
\label{sec:totxs}
Next, we consider the DIS structure functions at small Bjorken-$x$.
The overall normalization of the dipole amplitude depends on the strong coupling constant $\as = g^2/(4\pi)$, see Eq.~\eqref{eq:dipole_from_lcpt}. The same coupling constant also affects the Bjorken-$x$ dependence of the dipole amplitude via the BK evolution. In this work, our strategy is to fix the value of $\as$ by calculating the total charm production cross section, and comparing it to the HERA reduced cross section data from Ref.~\cite{H1:2018flt}. We set the initial condition for the BK evolution at $x_0=0.01$, and compare it to the HERA data in the region $x<0.01, Q^2<100\,\mathrm{GeV}^2$ (note that the smallest $Q^2$ bin in the data is $Q^2=2.5~$GeV$^2$). In this region, there are $N=39$ data points. The experimental data is reported as reduced cross section
\begin{equation}
    \sigma_r(x,y,Q^2) = F_2(x,Q^2)
    - \frac{y^2}{1 + (1-y)^2} F_L(x,Q^2).
    \end{equation}
Here $y=Q^2/(sx)$ is the inelasticity variable, not to be confused with the evolution rapidity. The proton structure functions $F_2$ and $F_L$ are expressed in terms of the total cross section for the virtual photon-proton cross section $\sigma^{\gamma^*p}$:

\begin{align}
F_2(x,Q^2) &= \frac{Q^2}{4\pi \aem}\left(\sigma^{\gamma^* A}_{T} + \sigma^{\gamma^* A}_{L} \right), \\
F_L(x,Q^2) &= \frac{Q^2}{4\pi \aem}\sigma^{\gamma^* A}_{L}. 
\end{align}
In the dipole picture, the total cross section for the virtual photon-proton scattering can be expressed as a convolution of the photon wave function and the dipole amplitude as~\cite{Kovchegov:2012mbw}
\begin{equation}
\label{eq:cgc-gammap}
\sigma^{\gamma^* A}_{T,L} = 2\sum_f \int \dd[2]{\bt} \dd[2]{\rt} \dd{z} \, \left|\Psi^{\gamma^* \to q\bar q}(\rt,z,Q^2)\right|^2 \, N(\rt,\bt,\bar x).
\end{equation}
Here $f$ is the quark flavor, $Q^2$ the photon virtuality and $\Psi^{\gamma^* \to q\bar q}$ is the leading order wave function
for the $q\bar q$ Fock state of the virtual photon. In this equation we replace $N(\rt,\bt,\bar x)$ by the impact
parameter averaged dipole amplitude $N(r,\bar x)$, as described above, and $\int \dd[2]{\bt} \rightarrow S_T$.
We fix the mass of the $c$ quark to $1.4$~GeV. 
The dipole amplitude in Eq.~\eqref{eq:cgc-gammap} is evaluated at $\bar x = x(1+4m_q^2/Q^2)$, where $m_q$ is the quark mass which enforces a smooth approach to the photoproduction limit~\cite{Golec-Biernat:1998zce,Albacete:2010sy}.

 \begin{figure}
     \centering
     \includegraphics[width=\columnwidth]{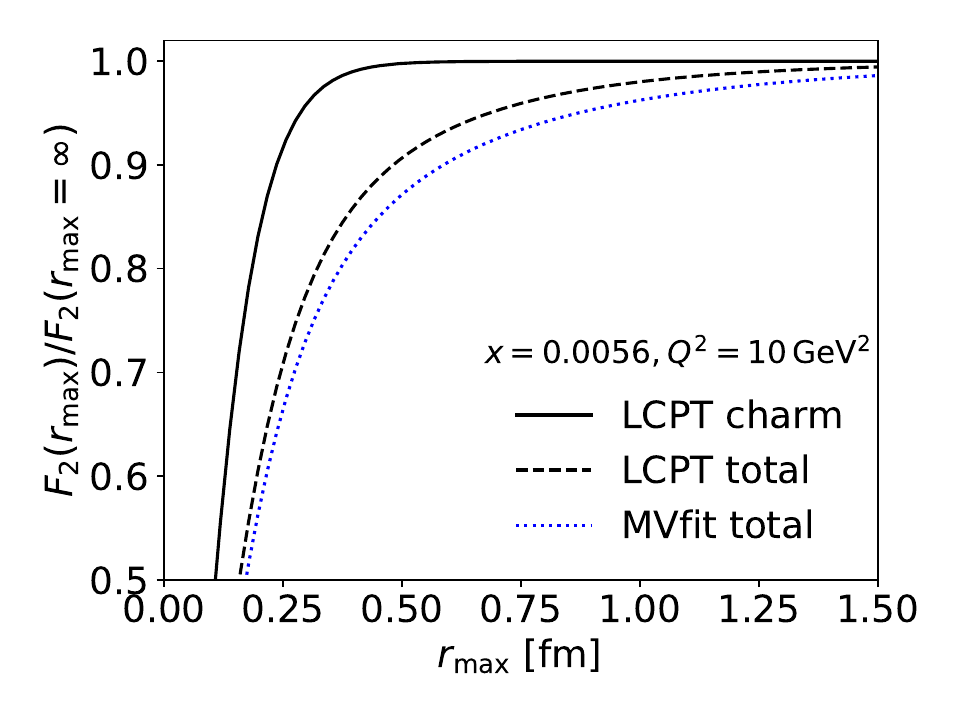}

     \caption{The fraction of the charm and inclusive DIS structure functions at $x=0.0056, Q^2=10$~GeV$^2$ (corresponding to $\bar x=0.01$ in the case of charm production) as function of the cutoff on the dipole size
     imposed in Eq.~\eqref{eq:cgc-gammap}. The total cross section is the sum of light quark (mass $0.14\,\mathrm{GeV}$) and charm quark (mass $1.4\,\mathrm{GeV}$) production contributions.
   }
     \label{fig:maxrdep}
 \end{figure}
 In order to confirm that the charm production cross section is not sensitive to non-perturbatively large dipoles we show  in fig.~\ref{fig:maxrdep} the fraction of the total cross section at $x=0.0056, Q^2=10\,\mathrm{GeV}^2$ as a function of the upper limit 
 $r_\text{max}$ for the $r$ integral in
Eq.~\eqref{eq:cgc-gammap}. It is evident that the charm cross section is saturated by small dipoles whereas the inclusive cross  section (calculated using $m_q=0.14~$GeV for the light quarks)
 at $Q^2=10$~GeV$^2$ is sensitive to larger dipoles  beyond sizes where we may trust our calculation. When using a modified MV-model parameterization as an initial condition for the evolution with different extrapolation in the infrared region, one needs to integrate up to even larger $r$ in order to recover the full result for $F_2$. The charm production cross section is not affected by the different large-$r$ extrapolation (not shown).
Qualitatively similar results have been obtained with the commonly used IPsat parameterization for the dipole-proton amplitude where, typically, even larger dipole sizes  contribute as compared to the setup with factorized impact parameter dependence applied here~\cite{Mantysaari:2018nng,Kowalski:2003hm}.
For these reasons we shall focus on charm production. In the future, our approach could be applied to other hard, perturbative
processes such as, for example, single-inclusive particle production at high enough transverse momentum.

\begin{figure}
    \centering
    \includegraphics[width=\columnwidth]{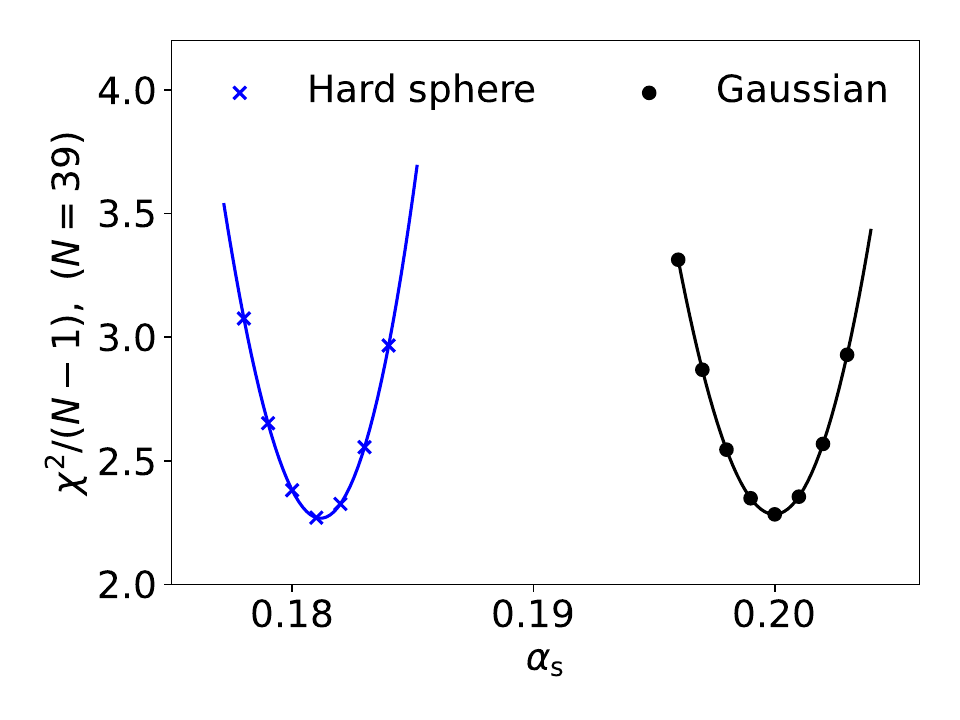}
     \caption{Determining the strong coupling constant by fitting the HERA charm production data in the region $Q^2<100\,\mathrm{GeV}^2, x<0.01$, with leading-log BK evolution started at $x_0=0.01$. The solid lines are polynomial fits to the computed values used to extract the optimal values for $\as$. 
     The optimal values are $\as=0.200, \chi^2/(N-1)=2.27$ for the Gaussian proton and $\as=0.181, \chi^2/(N-1)=2.28$ for the hard sphere proton.
     }
    \label{fig:alphas_fit}
\end{figure}

Considering only the strong coupling constant $\as$ as free parameter we obtain a reasonably good description of the charm production data.
The 
value of $\chi^2/(N-1)$ as a function of $\as$ is shown in Fig.~\ref{fig:alphas_fit} using two different density profiles (Gaussian and hard sphere) for the proton. These two setups have different upper limits for the impact parameter $b$ and correspondingly different transverse areas for the proton. The extracted optimal values for the strong coupling constant are $\as=0.200$ for the Gaussian proton and $\as=0.181$ for the hard sphere profile.  These values are used throughout this paper.
We note that fits of the total (rather than charm) cross section with tuned initial conditions~\cite{Albacete:2010sy,Lappi:2013zma} 
and running coupling corrections to the BK equation have achieved lower
$\chi^2/N \approx 1$, without being able to simultaneously describe the charm data~\cite{Albacete:2010sy}. However, with our {\em calculated} initial condition there is room for the expected improvements of
the photon wave function, evolution equation, and, of course, of the initial condition.

In this analysis, we have fixed the collinear regulator to $m_\text{coll} = 0.2\,\mathrm{GeV}$ in the LCPT calculation of the initial condition. As the charm cross section is dominated by small dipoles, our results are not highly sensitive to the actual value of this regulator: changing $m_\text{coll}$ by a factor of $2$ changes $\chi^2/(N-1)$ to HERA data by only $2\dots 5\%$ when using the optimal $\as$. 
We also keep the charm mass fixed to $1.4~\mathrm{GeV}$. The optimal value for $\as$ and the achieved $\chi^2/(N-1)$ naturally will depend on this choice. 
We choose to work with fixed quark mass and collinear regulator and do not attempt to fit these simultaneously with $\as$, as the purpose of this work is to demonstrate the feasibility of \emph{computing} the initial condition for the BK equation, and we emphasize that numerically potentially important higher order effects such as running coupling are still missing from the setup.

\begin{figure*}[t]
    {
    \includegraphics[width=0.32\textwidth]{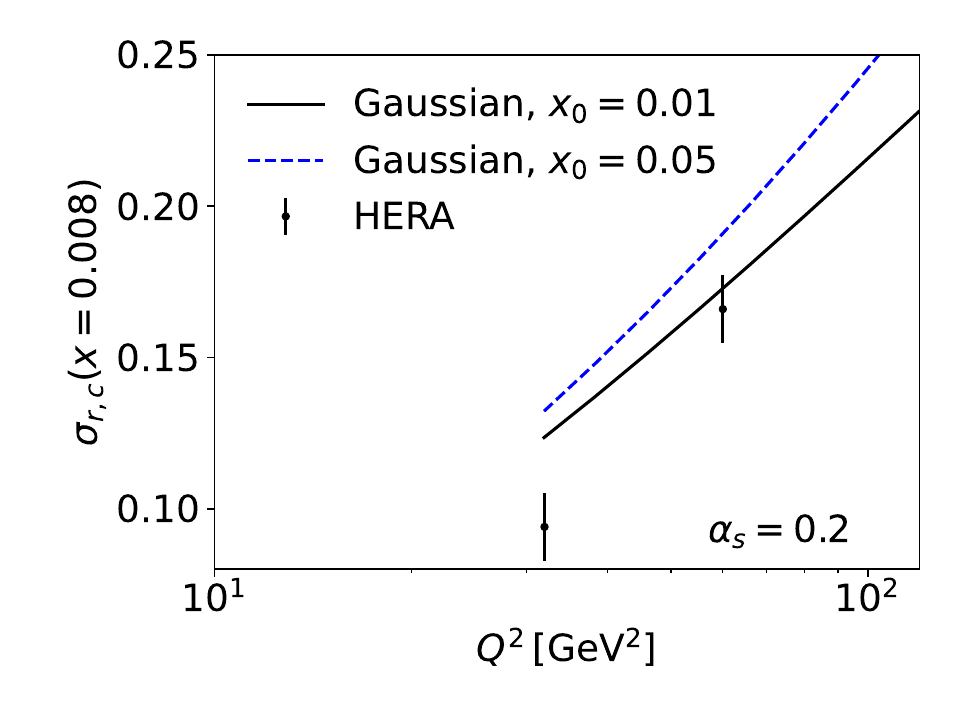}
    \label{fig:charm_x_0.008}
    }
    {
    \centering
    \includegraphics[width=0.32\textwidth]{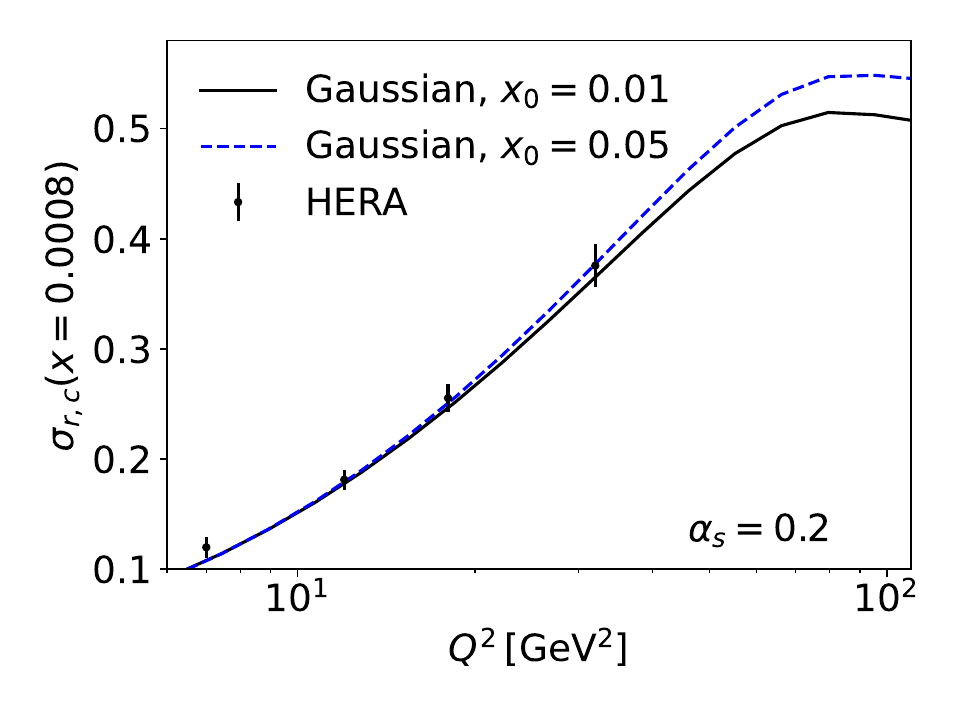}
     \label{fig:charm_x_0.0008}
    }
    {
    \centering
    \includegraphics[width=0.32\textwidth]{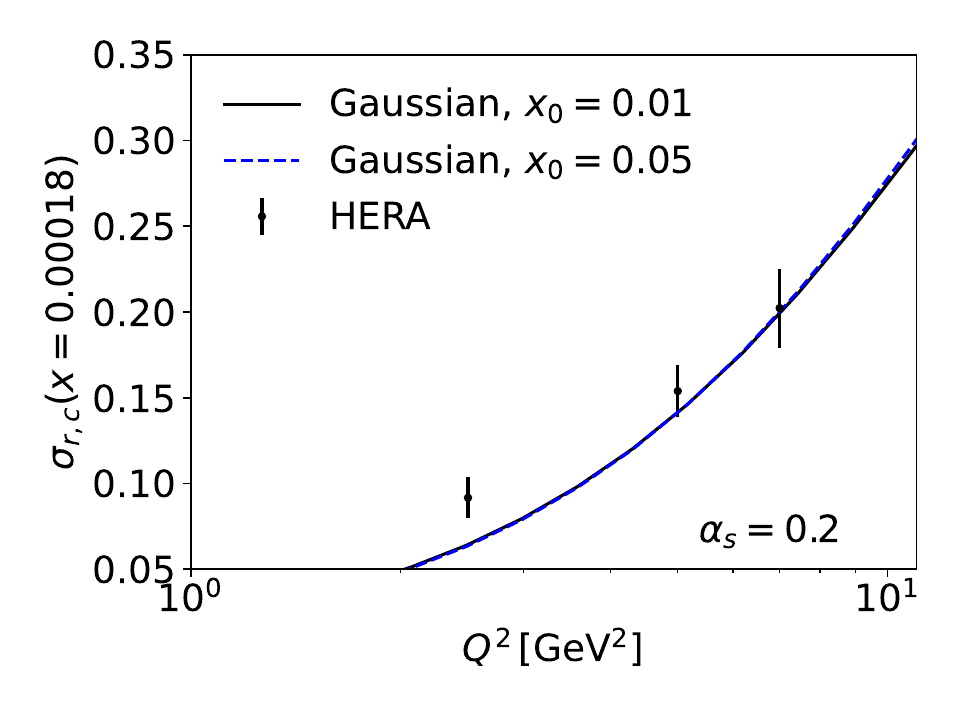}
     \label{fig:charm_x_0.00018}
    }
    \caption{Charm production reduced cross section at $\sqrt{s}=318\,\mathrm{GeV}$ compared to the HERA data. Results are shown in the region where $\bar x \le 0.01$. }
     \label{fig:charm_q2dep}
\end{figure*}

A comparison to the HERA charm production data in different Bjorken-$x$ bins is shown in Fig.~\ref{fig:charm_q2dep} as a function of the photon virtuality. We have checked that these results remain the same if we use
the analytic parameterization~\eqref{eq:mv}, with parameters from table~\ref{tab:mvfit}, as initial condition; this confirms
the insensitivity of the charm cross section to the large-$r$ extrapolation of the scattering amplitude.

At $x=0.008$ there is only very little ($\le 0.2$ units of rapidity when $x_0=0.01$) evolution, so the dipole amplitude is almost completely determined by our initial condition. On the other hand, we also show results at lower $x$ which is dominated by the BK evolution. In addition to our standard setup where the initial condition for the BK evolution is set at $x_0=0.01$, we also show results using an initial condition computed at larger  $x_0=0.05$. Note the weak dependence of this observable, at least, on where the ``hand-off'' from the $x$-dependent
initial condition to the evolution equation occurs. In contrast, ad-hoc initial condition parametrizations have to be re-tuned when $x_0$ is changed.

Fig.~\ref{fig:charm_q2dep} shows a fair agreement of the reduced cross section obtained from our light-cone wave function
with the HERA charm data.
Close to the initial condition we obtain a slightly slower $Q^2$ dependence than seen in the data. As a result of the evolution this changes into faster virtuality dependence at very small $x$. This is because the BK evolution at fixed coupling develops a small anomalous dimension $\gamma\approx 0.6$  for the dipole amplitude and a smaller anomalous dimension results in faster $Q^2$ dependence.

 \begin{figure}
     \centering
     \includegraphics[width=\columnwidth]{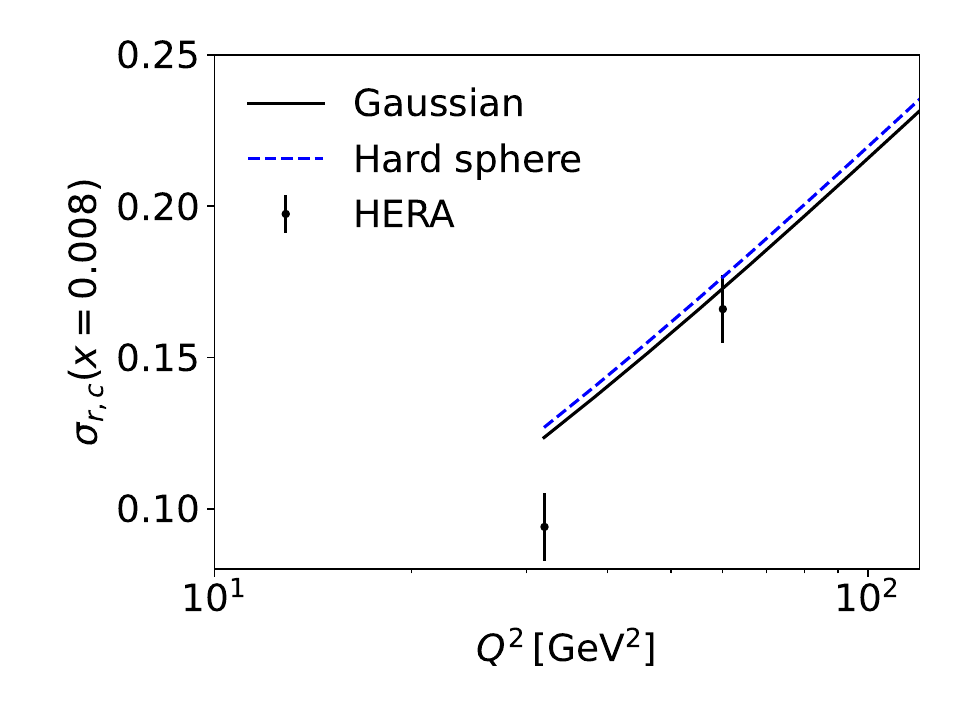}
     \caption{Reduced cross section at $\sqrt{s}=318\,\mathrm{GeV}$ calculated close to the initial condition using the Gaussian and Hard sphere density profiles, respectively.
   }
     \label{fig:gaussian_hard_sphere_sigmar}
 \end{figure}
 
Lastly, in Fig.~\ref{fig:gaussian_hard_sphere_sigmar} 
we study how sensitive the charm production cross section is on the chosen proton density profile, and as such on the maximum impact parameter $b_\mathrm{max}$ used in Eq.~\eqref{eq:ic_bavg}. The cross section is calculated at $x=0.008$ which is close to the initial condition for the BK evolution, again set at $x_0=0.01$. In both cases, we use the optimal value for the strong coupling constant extracted above.
The cross section increases only
slightly when the hard sphere profile with larger $b_\text{max}$ is used, but the dependence on the virtuality is not affected. This weak dependence on the selected proton profile confirms that our results are not sensitive to non-perturbatively large impact parameters.

\section{Summary and discussion}

The present work represents a first attempt at relating the short-distance structure of the proton at high energies to its low-energy properties, covering several
orders of magnitude in energy. We start from an effective three quark light-cone wave function which models the non-perturbative longitudinal and transverse momentum
distributions of quarks at $x\gsim 0.1$ as well as some of their correlations. The next step involves the computation, using exact kinematics, in light-cone perturbation theory
of the ${\cal O}(g)$ correction to the
light-cone wave function due to the emission of a gluon, and the ${\cal O}(g^2)$ virtual corrections due to the exchange of a gluon by two quarks. 
This provides a leading twist contribution to the dipole scattering amplitude, $N(r)\sim r^{2\gamma}$ with $\gamma=1$ at small $r$.
Optimistically, the LCPT correction
extends the validity of the resulting light-cone wave function into the regime of perturbative transverse momenta,
and parton momentum fractions $x\gsim 0.01$. The corresponding dipole scattering amplitude is then evolved to yet higher energies (lower
$x$) by solving the
BK equation, which resums emissions of additional soft gluons, and generates power corrections and an anomalous dimension.

The convolution of the LO photon light-cone wave function with the impact parameter averaged BK dipole scattering amplitude at leading logarithmic accuracy provides a fair
description of the reduced DIS charm cross section measured at HERA, for $\as \simeq 0.2$.
This value of the strong coupling was obtained from a fit to the charm cross section at $Q^2<100$~GeV$^2$ and $x<0.01$. None of the parameters of the low-energy three-quark
model wave function were re-tuned to the high-energy data. Despite the fair description of the highly accurate data the resulting $\chi^2/N_\text{dof} \approx 2.27$
with $N_\text{dof} = 38$
implies a very low statistical significance, i.e.\ a very low probability that the data represents statistical fluctuations about the model 
predictions: the integral over the $\chi^2$ distribution from $\chi^2=2.27\times 38$ to infinity, commonly denoted
as the "p-value", is $1.3\times 10^{-5}$. However, the very low statistical significance of the fit should not be confused with a
need for large corrections, Fig.~\ref{fig:charm_q2dep} shows that this is clearly not the case. 
This is entirely expected since there are multiple {\em known} sources of corrections such as, for example, of the photon wave function~\cite{Beuf:2021srj,Beuf:2021qqa,Beuf:2022ndu}, of the evolution
equation~\cite{Lappi:2016fmu,Balitsky:2008zza,Iancu:2015vea,Ducloue:2019ezk,Iancu:2015joa}, and, of course, of the initial condition for the evolution equation (our proton light-cone wave function) which, e.g., may be improved with running coupling corrections.
The data requires fairly moderate but {\em systematic} improvements of the model predictions across the relevant
ranges of $x$ and $Q^2$.

We have also provided analytic parameterizations of the impact parameter averaged dipole scattering amplitude for $x=0.01 \dots 0.05$ which accurately fit the numerical
data in the regime of perturbative dipoles, $r\lesssim 0.5$~fm. These parameterizations can be used in practice to estimate the corrections predicted by more accurate
evolution equations.
In the supplementary material we also provide the tabulated numerical data for $N(r,x)$. Their large-$r$ extrapolation differs from that of the
analytic parameterizations which allows for tests of the (in-)sensitivity to the uncontrolled non-perturbative regime of large dipoles.
The quest for more accurate theoretical predictions at high energy for the upcoming EIC at BNL~\cite{AbdulKhalek:2021gbh} and the proposed LHeC/FCC-he at CERN~\cite{LHeC:2020van} requires initial conditions for the evolution equations
which do not absorb theoretical improvements into a re-tune of their parameters.

\subsection*{Acknowledgments}
This material is based upon work supported by the U.S.\ Department of Energy, Office of Science, Office of Nuclear Physics, within the framework of the Saturated Glue (SURGE) Topical Theory Collaboration.
A.D.\ acknowledges support by the DOE Office of Nuclear Physics through Grant DE-SC0002307, and The City University of New York for PSC-CUNY Research grant 65079-00~53. 
This work was supported by the Academy of Finland, the Centre of Excellence in Quark Matter and projects 338263 and 346567 (H.M), and projects 347499 and 353772 (R.P). H.M is also supported under the European Union’s Horizon 2020 research and innovation programme by the European Research Council (ERC, grant agreement No. ERC-2018-ADG-835105 YoctoLHC) and by the STRONG-2020 project (grant agreement No. 824093). The content of this article does not reflect the official opinion of the European Union and responsibility for the information and views expressed therein lies entirely with the authors. Computing resources from CSC – IT Center for Science in Espoo, Finland and from the Finnish Grid and Cloud Infrastructure (persistent identifier \texttt{urn:nbn:fi:research-infras-2016072533}) were used in this work.

\bibliography{refs}
\bibliographystyle{JHEP-2modlong}

 \end{document}